\documentclass{article}
\usepackage{blindtext, graphicx}

\usepackage[margin=1in]{geometry}

\usepackage{cite}

\usepackage[cmex10]{amsmath}

\usepackage{hyperref}
\usepackage{enumerate}

\usepackage{listings}
\usepackage{color}

\usepackage{titling}

\usepackage[utf8]{inputenc}
 
\definecolor{codegreen}{rgb}{0,0.6,0}
\definecolor{codegray}{rgb}{0.5,0.5,0.5}
\definecolor{codepurple}{rgb}{0.58,0,0.82}
\definecolor{backcolour}{rgb}{0.95,0.95,0.92}
 
\lstdefinestyle{mystyle}{
    backgroundcolor=\color{backcolour},   
    commentstyle=\color{codegreen},
    keywordstyle=\color{magenta},
    numberstyle=\tiny\color{codegray},
    stringstyle=\color{codepurple},
    basicstyle=\footnotesize,
    breakatwhitespace=false,         
    breaklines=true,                 
    captionpos=b,                    
    keepspaces=true,                 
    numbers=left,                    
    numbersep=5pt,                  
    showspaces=false,                
    showstringspaces=false,
    showtabs=false,                  
    tabsize=2
}
 
\author{
Andrew Elkouri \\
Dept. of Mathematics, Statistics and Physics \\
Wichita State University \\
\href{mailto:axelkouri@wichita.edu}{\texttt{axelkouri@wichita.edu}}}

\title{Predicting the Sentiment Polarity and Rating of Yelp Reviews}

\date{\vspace{-5ex}}

\begin{document}

\maketitle

\section{Abstract}
Online reviews of businesses have become increasingly important in recent years, as customers and even competitors use them to judge the quality of a business. Yelp is one of the most popular websites for users to write such reviews, and it would be useful for them to be able to predict the sentiment or even the star rating of a review. In this paper, we develop two classifiers to perform positive/negative classification and 5-star classification. We use Naive Bayes, Support Vector Machines, and Logistic Regression as models, and achieved the best accuracy with Logistic Regression: 92.90\% for positive/negative classification, and 63.92\% for 5-star classification. These results demonstrate the quality of the Logistic Regression model using only the text of the review, yet there is a promising opportunity for improvement with more data, more features, and perhaps different models.

\section{Introduction}
\subsection{Overview}
As customers and competitors rely on Yelp reviews to judge the quality of a business, it is important for Yelp to be able to predict the sentiment polarity and rating of a given review. With Yelp's newly released dataset \cite{yelp-dataset}, we perform two types of classifications based on the review text alone: simple positive/negative classification, and a star rating (1 through 5 inclusive). To build these classifiers, we will use Naive Bayes, Support Vector Machines, and Logistic Regression. Note that we are using the review text as the only input to the classifier (e.g., given the review ``The best British food in New York'', we want to predict `positive', or 5 stars).
\subsection{Motivation}
\indent It is useful for Yelp to associate review text with a star rating (or at least a positive or negative assignment) accurately in order to judge how helpful and reliable certain reviews are. Perhaps users could give a good review but a bad rating, or vice versa. Also Yelp might be interested in automating the rating process, so that all users would have to do is write the review, and Yelp could give a suggested rating.

\section{Problem Definition}
\subsection{Overview}
Given a review of a business, we try to solve two problems: positive or negative (sentiment polarity) classification, and 5-star classification. In both cases the input to the classifier is a string representing a review, and the output (or class label) is either \textit{positive} or \textit{negative} in the former case, and an integer in the interval $[1, 5]$ in the latter case.
\subsection{Preprocessing, Vectorizer, Transformer, Classifier}
\indent The input to the classifier is text only, but this text will be uncleaned with a lot of unnecessary characters. For example, it could have unusual capitalization, extra whitespace, and of course punctuation, which is not needed for classification. Therefore all of these things must be cleaned. Furthermore, we are only interested in tokens with two or more alphanumeric characters. Some tokens may be stop words, that is, words that convey no information but are very common in English. So we should remove these also, as they will reduce the accuracy of the classifier (or at least slow down the classifier as there would be more unnecessary text to process). \\
\indent Once the text is cleaned, we need to build feature vectors. We will use a bag of words representation to store the occurrences of each word in a matrix. (See the next section for more details.) Now that we have these occurrence counts, we still need to weigh each word, i.e., how important it is to a review in the corpus. We will use the \textit{tf-idf} statistic for this. Finally we can build the classifier with either Naive Bayes, Support Vector Machines, or Logistic Regression.

\section{Solution Technique}
There are four steps to perform: first we preprocess the data, then we build a vectorizer, then a transformer, and finally the classifier. For all of these steps we use the \textit{scikit-learn} package for Python.
\subsection{Preprocessing the Data}
Text preprocessing is done by a \texttt{CountVectorizer} class from \texttt{sklearn.feature\_extraction.text}. By default, a built-in regular expression specifies what is considered a token: it must have two or more alphanumeric characters (punctuation is always ignored). Furthermore, all the text is made lowercase with extra whitespace ignored. Stop words are also removed.
\subsection{Feature Vectors}
After preprocessing, we need to transform the text into feature vectors. For that we use the bag of words representation. To each word in all reviews we assign a unique integer ID. This assignment is done with a dictionary or map from words to integers. And for each review $\#i$, we count the number of occurrences of each word $w$, and this count is stored in $X[i, j]$, where $j$ is the index of $w$ in the dictionary. Note that the majority of the elements in $X$ will be zero, that is, $X$ is a sparse matrix. This is because for each review, a relatively small number of distinct words are used. (An implementation detail: to save memory, $X$ is stored as a \texttt{scipy.sparse} matrix, so that only the nonzero elements of the feature vectors remain in memory.)
\subsection{tf-idf Transformer}
Now we need to build the transformer to calculate the weight of each word. For this we use the \textit{tf-idf} statistic. So at this point, although we have a count of the occurrences of each word, there is a discrepancy between long and short reviews, as the former will have a higher count than the latter. Therefore we divide the occurrence count of each word in a review by the total number of words in the review. We call these new features Term Frequencies, or \textit{tf}. \\
\indent There remains a problem with \textit{tf}: some words could appear in many reviews ranging from 1 to 5 stars, so they are not informative because they are not unique to a certain class of reviews. The solution is called Inverse Document Frequency, or \textit{idf}. The idea is to offset or downscale the weight of a word based on its frequency in the corpus.
\subsection{Class Labels}
Finally, the class labels in this problem are simple. In the case of binary positive or negative classification, a review is assigned the \textit{positive} label if its star rating is greater than or equal to 3, and any review with a rating less than 3 is assigned the \textit{negative} label. In the case of 5-star classification, the review is assigned its star rating.
\subsection{Building the Classifier}
To build the classifier we use three different supervised techniques: Naive Bayes, Support Vector Machines, and Logistic Regression. These techniques were chosen as they are simple to understand and implement, they run relatively quickly, and they have historically given good results for text classification \cite{prediction-yelp-stars}. In each case we used 70\% of the data for training, and the rest for testing.
\subsection{Implementation}
To read the data file, which has a JSON object on each line, we use a function \texttt{loadData(filename, startLine, endLine)} to read between lines \#\texttt{startLine} and \#\texttt{endLine}. \texttt{loadData} returns two lists: one containing each review, and the other containing the corresponding rating. The elements of the latter will be either \textit{positive} or \textit{negative}, or an integer in the interval $[1, 5]$, depending on whether we do positive/negative or 5-star classification. \\
\indent Note that we use \texttt{loadData} to load both the training and testing data. So of course, the intersection between the two sets is empty, thanks to the interval of line numbers as parameters. \\
\indent Now that the data is loaded and we have built the vectorizer and transformer, we can build the classifier using \texttt{sklearn.pipeline.Pipeline}. The \texttt{Pipeline} class ``behaves like a compound classifier'' \cite{scikit-pipeline} class. The transformer is the \textit{tf-idf} statistic, and the classifier is built using either \texttt{sklearn.naive\_bayes.MultinomialNB} for Naive Bayes, \texttt{sklearn.linear\_model.SGDClassifier} for Support Vector Machines, or, for Logistic Regression, \texttt{sklearn.linear\_model.LogisticRegression}.
\indent For example, using the \texttt{Pipeline} class, we can build and train a Naive Bayes classifier as in the code in Figure ~\ref{fig:code}.
\begin{figure}
\begin{lstlisting}[basicstyle=\ttfamily, language=Python]
NB_clf = Pipeline([('vect', CountVectorizer(stop_words='english')),
                   ('tfidf', TfidfTransformer()),
                   ('clf', MultinomialNB()),
        ])

NB_clf = NB_clf.fit(train_data, train_labels)
\end{lstlisting}
\caption{Building and training a Naive Bayes Classifier}
\label{fig:code}
\end{figure}

In Figure ~\ref{fig:code}, we first instantiate \texttt{CountVectorizer} and specify that we want to remove English stop words. Then we use \texttt{TfidfTransformer} to calculate the weights of each word, and lastly we create the Naive Bayes classifier with \texttt{MultinomialNB}. After the classifier is built, we train it by simply passing in the training data and training labels.


\section{Dataset}
For this project we use the dataset from the Yelp Dataset Challenge. The data are in JSON format. The format of the review data is shown in Figure ~\ref{fig:data}. We are interested in two fields only: `text' and `stars'. The other fields are not used here, but they could be useful as features for future work (see the final Discussion section). \\
\begin{figure}
\begin{lstlisting}[basicstyle=\ttfamily]
{
    'type': 'review',
    'business_id': (encrypted business id),
    'user_id': (encrypted user id),
    'stars': (star rating, rounded to half-stars),
    'text': (review text),
    'date': (date, formatted like '2012-03-14'),
    'votes': {(vote type): (count)},
}
\end{lstlisting}
\caption{Example JSON object from Yelp's review data}
\label{fig:data}
\end{figure}
\indent The file containing the review data has a JSON object formatted as in Figure ~\ref{fig:data} on each line. This file has 1,569,264 reviews, of which we use varying amounts (see Figures ~\ref{fig:pndataplot} and ~\ref{fig:5dataplot}). The review text in the data file is uncleaned and taken directly from the website without modification. Therefore, as described above, some text preprocessing was necessary before building the classifier. To read this data we use the \texttt{json} package built in to Python. \\
\indent It is worth mentioning that this review data is by no means comprehensive. It includes data from 10 cities total, in the United Kingdom, Germany, Canada, and the United States. Furthermore, the data comes from all types of businesses on Yelp, not just restaurants, for instance.

\section{Results}

The results for both positive/negative classification and 5-star classification are shown in Figure ~\ref{fig:accuracy}.

\begin{figure}
\begin{center}
\begin{tabular}{ |c|c|c|c| }
\hline
\textbf{Learning Model} & \textbf{Positive/Negative Test Accuracy} & \textbf{5-star Test Accuracy}\\
\hline
Naive Bayes & 87.19\% & 55.77\% \\
\hline
SVM & 83.06\% & 59.56\% \\
\hline
LR & 92.90\% & 63.92\% \\
\hline
\end{tabular}
\end{center}
\caption{Test accuracy of Naive Bayes and SVM}
\label{fig:accuracy}
\end{figure}

\subsection{Positive/Negative Classification}
\begin{figure}
\begin{center}
\includegraphics[scale=0.22]{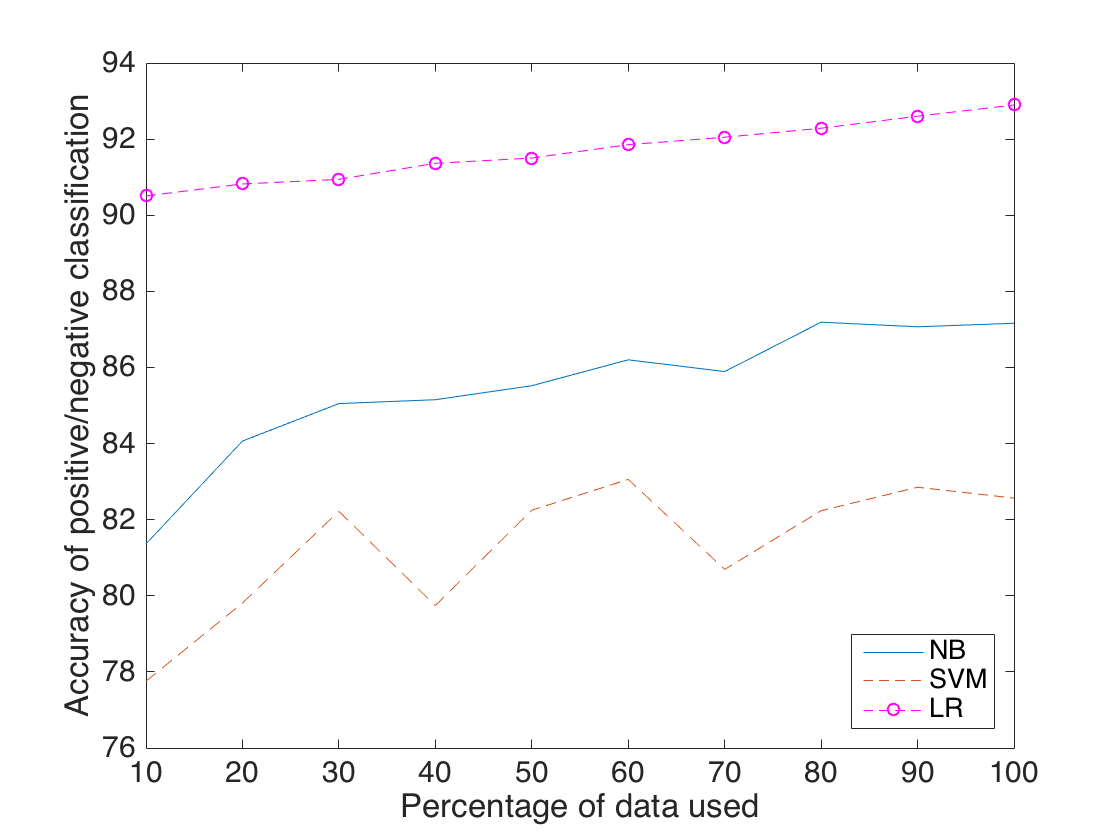}
\end{center}
\caption{Percentage of Data Used vs Accuracy of Positive/Negative Classification}
\label{fig:pndataplot}
\end{figure}

The positive/negative classification results are fairly high, partly because it is a rather simple problem, at least compared to 5-star classification. Logistic Regression performed the best by 5.71\% compared to Naive Bayes, which outperformed Support Vector Machines by 4.13\%. \\
\indent We also measured the relationship between the amount of data used for training and the accuracy of the classifier, which is displayed in Figure ~\ref{fig:pndataplot}. It is immediately obvious that increasing the amount of data helps only to an extent, and it depends heavily on the model used. The benefit of more data is clearer in the case of Naive Bayes and Logistic Regression. The accuracy of Support Vector Machines fluctuated quite a bit with changes in the amount of data used, which could be a result of overfitting.

\subsection{5-star Classification}
For 5-star classification, Support Vector Machines outperformed Naive Bayes in this more complicated task. However Logistic Regression still did the best.

\begin{figure}
\begin{center}
\includegraphics[scale=0.22]{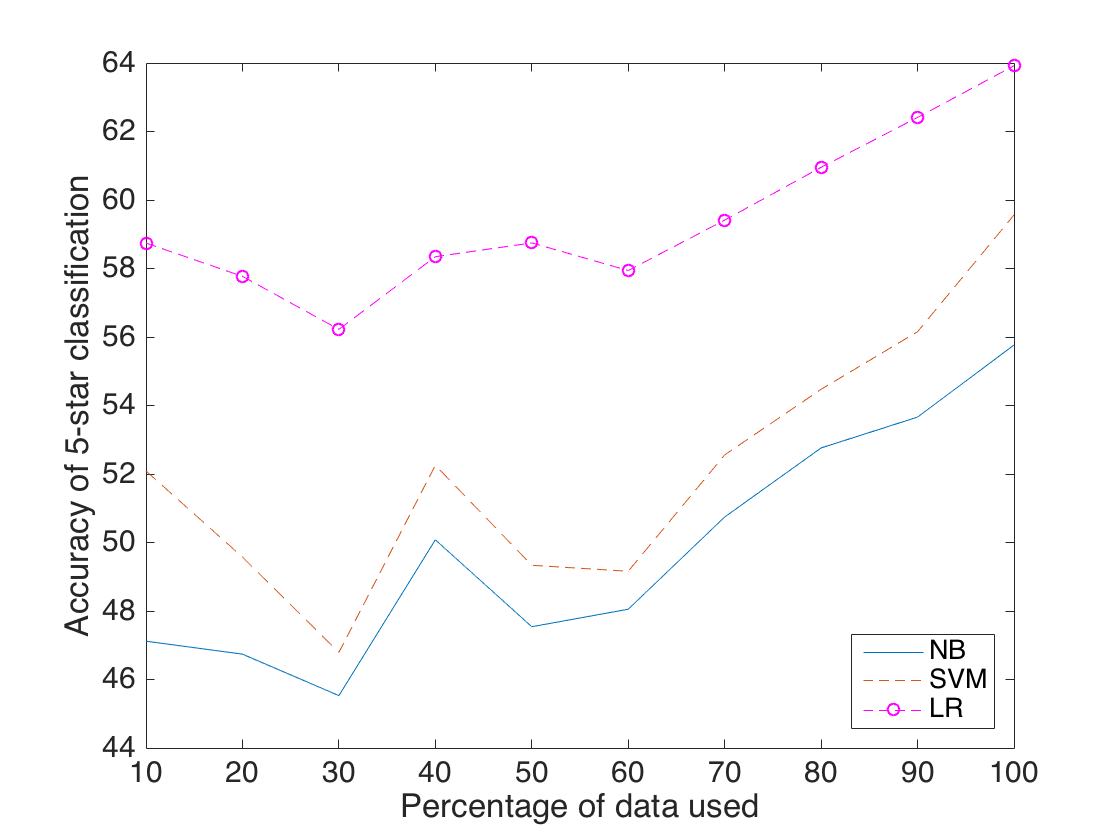}
\end{center}
\caption{Percentage of Data Used vs Accuracy of 5-star Classification}
\label{fig:5dataplot}
\end{figure}

\indent As with positive/negative classification, we examined the relationship between the amount of data used and 5-star classification accuracy, which is displayed in Figure ~\ref{fig:5dataplot}. In this case, the accuracies of both Naive Bayes and Support Vector Machines seemed to change by similar amounts in the same places. The accuracy of Logistic Regression also declined at 30\% and 60\% data usage, but thereafter increased considerably.

\subsection{Comparison of Models}
\subsubsection{Speed}
In Figure ~\ref{fig:time}, we display the time (in seconds) taken by the three models using 100\% of the data. Naive Bayes and Support Vector Machines are quite similar in their running times, and Logistic Regression is by far the slowest of the three, particularly in the case of 5-star classification. However, depending on the amount of data, using Logistic Regression could still be a worthwhile trade-off considering its higher accuracy.

\begin{figure}
\begin{center}
\begin{tabular}{ |c|c|c|c| }
\hline
\textbf{Learning Model} & \textbf{Positive/Negative Duration} & \textbf{5-star Duration} \\
\hline
Naive Bayes & 178.52 & 176.03 \\
\hline
SVM & 174.41 & 178.02 \\
\hline
LR & 277.66 & 585.92 \\
\hline
\end{tabular}
\end{center}
\caption{Time (in seconds) of Naive Bayes, SVM, and LR using 100\% of the data}
\label{fig:time}
\end{figure}

\subsubsection{Accuracy}
Like their speeds, the accuracies for both Naive Bayes and Support Vector Machines are quite similar, differing by at most about 4\% when using the same amount of data. The accuracy of Logistic Regression is the highest in both types of classification.

\section{Discussion}
\subsection{Overview of Results}
The results we obtained are similar to what we had anticipated in some ways, and surprising in other ways. First of all, positive/negative classification is a much simpler problem than 5-star classification, largely because it is less specific; for example, both 1 and 2 star reviews will be classified as 'negative', whereas with 5-star classification, it could be hard to tell the difference between the reviews since they will both use similar language. Therefore it was expected that we would have far better accuracy for positive/negative classification than for 5-star classification, which was indeed the case. Another expected result was that more data would improve accuracy. This turned out to be true to some extent, as more data provided further training examples. \\
\indent One surprising result was the better performance of Naive Bayes in the case of positive/negative classification. It is a very simple classifier, and we had expected Support Vector Machines to outperform it in both types of classification. The difference in accuracy in both cases is about 4\%, which is slight but not insignificant. \\
\indent Another surprise was the excellent performance of Logistic Regression. It was expected to outperform Naive Bayes, but not necessarily Support Vector Machines, and not by such a margin: its accuracy was 9.84\% and 4.36\% higher than Support Vector Machines in positive/negative and 5-star classification respectively.

\subsection{Comparisons to Other Results}
According to Jong's paper ``Predicting Rating with Sentiment Analysis'' \cite{predict-sa-jong}, which also uses Yelp's dataset, ``for opinionated texts, there is usually a 70\% agreement between human raters''. Our best result for positive/negative classification exceeds that by 22.90\%. However, for 5-star classification, our maximum accuracy is about 6.08\% lower, but far better than a random star guess, which would have an accuracy of only 20\%. \\
\indent For positive/negative classification, Jong achieved a maximum test accuracy of about 78\% with Naive Bayes, compared to our maximum test accuracy of 92.90\% with Logistic Regression. \\
\indent Carbon et al. \cite{applications-ml-predict} worked on 5-star classification of Yelp reviews as well, using several models including Gaussian Discriminant Analysis and Logistic Regression. Using their entire feature set, they achieved a maximum test accuracy of 46.09\% with GDA, compared to our maximum test accuracy of 63.92\% with Logistic Regression.

\subsection{Future Work}
\begin{enumerate}[i.]
\item \textit{Use a different learning model.} The models used here are quite simple. A more advanced one such as Random Forests or Neural Networks could produce better results, perhaps with a trade-off in time depending on the model.
\item \textit{Use more data.} Although there are almost 1.6 million reviews in Yelp's dataset, more data could improve classification accuracy. In our results, the accuracy of Logistic Regression in particular increased with greater data usage (see Figures ~\ref{fig:pndataplot} and ~\ref{fig:5dataplot}).
\item \textit{Use more features.} The only input to our classifiers was the review text. However, there is an abundance of other interesting features in Yelp's dataset that could be used for classification. For example, each review itself has a rating (called `votes') of how funny, cool, or useful it is. Perhaps a review with many such votes could be considered more reliable than other reviews with fewer votes, and therefore the predicted rating from the classifier would be deemed more trustworthy with a higher score assigned to it. Another idea would be to use the business ID of the review to look up attributes of the business to help you determine the reliability of a review's rating. So you would effectively be building and combining two different classifiers (one for the review, and one for the business being reviewed), both with the goal of predicting a review's star rating.
\end{enumerate}

\subsection{Conclusion}
Overall the results were satisfactory for positive/negative classification, but there is room for improvement for 5-star classification. There are many general challenges to overcome besides using more data or more features for 5-star classification. As Pang and Lee discuss, ``some potential obstacles to accurate rating inference include lack of calibration (e.g., what an understated author intends as high praise may seem lukewarm), author inconsistency at assigning fine-grained ratings, and ratings not entirely supported by the text'' \cite{pang-lee-seeingstars}. These are nontrivial problems, and it is hard to imagine how to solve them by any kind of data preprocessing before building the feature vectors, or by other means. With that said, there are several promising avenues for future research which could considerably improve 5-star classification accuracy.

\end{document}